# Solitons of the coupled Schrödinger - Korteweg - de Vries system with arbitrary strengths of the nonlinearity and dispersion

**Gromov Evgeny**[1, a] **and Malomed Boris**[2,3]

[1]National Research University Higher School of Economics, Nizhny Novgorod 603155, Russia

[2]Department of Physical Electronics, School of Electrical Engineering, Faculty of Engineering, and Center for Light-Matter Interaction,  Tel Aviv University, Tel Aviv 69978, Israel

[3]ITMO University, St. Petersburg 197101, Russia

**Abstract**

New two-component soliton solutions of the coupled high-frequency (HF) – low-frequency (LF) system, based on Schrödinger - Korteweg - de Vries (KdV) system with the Zakharov's coupling, are obtained for arbitrary relative strengths of the nonlinearity and dispersion in the LF component. The complex HF field is governed by the linear Schrödinger equation with a potential generated by the real LF component, which, in turn, is governed by the KdV equation including the ponderomotive coupling term, representing the feedback of the HF field onto the LF component. First, we study the evolution of pulse-shaped pulses by means of direct simulations. In the case when the dispersion of the LF component is weak in comparison to its nonlinearity, the input gives rise to several solitons in which the HF component is much broader than its LF counterpart. In the opposite case, the system creates a single soliton with approximately equal widths of both components. Collisions between stable solitons are studied too, with a conclusion that the collisions are inelastic, with a greater soliton getting still stronger, and the smaller one suffering further attenuation. Robust intrinsic modes are excited in the colliding solitons. A new family of approximate analytical two-component soliton solutions with two free parameters is found for an arbitrary relative strength of the nonlinearity and dispersion of the LF component, assuming weak feedback of the HF field onto the LF component. Further, a one-parameter (non-generic) family of exact bright-soliton solutions, with mutually proportional HF and LF components, is produced too. Intrinsic dynamics of the two-component solitons, induced by a shift of their HF component against the LF one, is also studied, by means of numerical simulations, demonstrating excitation of a robust intrinsic mode. In addition to the above-mentioned results for LF-dominated two-component solitons, which always run in one (positive) velocities, we produce HF-dominated soliton complexes, which travel in the opposite (negative) direction. They are obtained in a numerical form, and by means of a quasi-adiabatic analytical approximation. The solutions with positive and negative velocities correspond, respectively, to super- and subsonic Davydov-Scott solitons.

In the course of 60 years of intensive theoretical and experimental studies, many types of solitons, i.e., self-trapped localized states supported in dispersive media by dynamical balance with nonlinear interactions, have been discovered. In this vast zoo of nonlinear modes, important species are symbiotic ones, which are supported by interactions of high-frequency (HF) and low-frequency waves. A classical example is known in plasma physics, in the form of the interaction of the HF Langmuir waves (rapid motion of electrons in the ionized gas with respect to slowly moving heavy ions) and LF ion-acoustic waves (sound propagating through the plasma in the form of density perturbations). Direct nonlinear interactions between the HF and LF modes via three-wave and four-wave mixing do not occur, as their characteristic frequencies and wavenumbers differ by several orders of magnitude. Nevertheless, the effective interaction is possible in the case when phase and group velocities the HF and LF waves take close values. In terms of plasma physics, spatially inhomogeneous distributions of the HF intensity give rise to the ponderomotive source which generates LF acoustic waves, and, in turn, density perturbations carried by the LF waves create an effective potential which may locally trap the HF waves. A paradigmatic model of the LF-HF interactions is provided by the Zakharov's system, which was derived, in the context of the Langmuir-acoustic interactions in plasma physics, about 50 years ago [9]. In the framework of this system, the propagation of the HF obeys the linear Schrödonger equation with an effective potential, represented by the local density of the acoustic perturbations, and the LH field is governed by the Korteweg – de Vries (KdV) equation, which contains the intrinsic third-order intrinsic dispersion and quadratic nonlinearity of the ion-acoustic waves, and the above-mentioned ponderomotive term, which couples the LF component to the HF one. The Zakharov's system and its varieties were recognized as universal models governing the HF-LF interactions in a large number of settings with underlying physics which may be completely different from plasmas, such as the HF intramolecular vibrations and LF acoustic excitation in the Davydov's model of the dynamics of long polymer molecules [18], or the HF surface waves and LF internal waves in geophysical hydrodynamics [15-17]. In all these realizations of the Zakharov's systems, soliton solutions play a profoundly important role. In this paper, we address solitons as solutions of the coupled system of KdV and linear Schrödinger equations. Previously, soliton solutions of this realization of the Zakharov's system were known in some specific situations, such as the one with equal nonlinearity and dispersion coefficients in the KdV component, which makes it possible to find exact Davydov-Scott (DS) solitons [18]. In this work, we report results of a systematic numerical and analytical investigations of solitons in a broad context. In particular, soliton modes in the coupled linear Schrödinger – KdV system fall into two large classes, LF-dominated and HF-dominated ones. Solitons of the former type always run in one (positive) direction, like their KdV counterparts, while the HF-dominated solitons travel in the opposite direction, which is demonstrated numerically and in an approximate analytical form, based on a quasi-adiabatic method. The solutions with opposite signs of the velocity actually correspond to supersonic and subsonic DS solitons [20].

---

[a] Author to whom correspondence should be addressed. Electronic mail: egromov@hse.ru

## 1. Introduction

Solitons, as robust self-trapped modes, are generated by the wave propagation in diverse dispersive nonlinear media, including surface waves (SWs) on deep and shallow water, internal waves (IWs) in stratified liquids, Langmuir and ion-sound waves in plasma, pulses and beams in nonlinear photonics, matter waves in Bose-Einstein condensates, electromagnetic waves in long Josephson junctions, spin waves in magnetically ordered materials, etc. [1-8]. In the framework of rigorous mathematical analysis, term "solitons" is usually reserved only for exact solutions of integrable models [1], but in physics literature solitary waves in nonintegrable systems are also commonly called solitons. We use this term below in the same general sense.

Dynamics of intense low-frequency (LF) waves, such as IWs in stratified liquids, SWs on shallow water and ion-sound waves in plasmas, is modeled by the unidirectional Korteweg - de Vries (KdV) equation. Soliton solutions of these equations originate from the balance of nonlinearity and dispersion of the LF waves.

In many physically relevant settings, HF waves are naturally coupled to the propagation of LF excitations. This is possible because, while the frequencies and wavenumbers of the wave modes of the two types are widely different, their phase and group velocities may be close, thus providing resonant enhancement of the coupling. A generic class of models of the HF-LF interaction is represented by the Zakharov's system, which includes the linear Schrödinger (LS) equation coupled to the KdV equation [9]. In this system, the HF field is governed by the Schrödinger equation, with the coupling provided by an effective potential term produced by the LF waves [10-12]. For HF Langmuir waves in plasmas, the potential term accounts for the variation of the plasma density caused by LF ion-sound waves, while for HF SWs in stratified fluids the potential is induced by the flow current on the surface of the fluid caused by the LF IWs under the surface. In this system, the LF component is generically governed by the unidirectional KdV equation (derivations can be found in Refs. [13-17]). In addition to the proper nonlinearity and dispersion of the LF waves, it includes a quadratic driving term induced by the HF waves, which accounts for the coupling to them.

Soliton solutions of this system were obtained, thus far, in the following particular cases: (i) for equal nonlinearity and dispersion coefficients of the LF wave ($\beta$ and $\gamma$, respectively, see section 2 below), the exact two-component Davydov-Scott (DS) soliton solution is available [18,19]; (ii) for the same case, $\beta = \gamma$, gray- and dark-soliton varieties of DS were found [20]; (iii) for the generic case of unequal nonlinearity and dispersion coefficients ($\beta \neq \gamma$) and a small amplitude of the HF field, two-component multihump solitons were found in an approximate form [10]; (iv) neglecting the feedback action of the HF component on the LF wave, which induces a potential well in the HF equation, localized linear stationary states of the HF field were found too [15]. In the framework of general Zakharov-type model, which couples the nonlinear Schrödinger (NLS) equation for intense HF waves to the Boussinesq or KdV equation for the LF component through quadratic terms, one-parameter families of exact two-component solitons for interacting HF and LF waves were recently found too [21] (for systems with the LF equation of the Boussinesq, other solitons were earlier reported in Refs. [22] and [23]).

The above-mentioned solutions were obtained under very special conditions. For underlying physical models, such as the Zakharov's system for the plasmas [9], it is relevant to find more general soliton solutions for an arbitrary relative strength of the nonlinearity and dispersion of the LF waves. In this work, such solutions are found in the framework of the coupled LS and KdV equations. The system is formulated in Section 2, which also includes its Lagrangian and Hamiltonian representations. Numerical results are reported in Section 3. First, the evolution of a pulse-shaped input is studied by means of systematic simulations in subsection 3.1. For the case of weak dispersion, in comparison with the nonlinearity ($\gamma << \beta$), the initial pulse splits into several two-component solitons, each having the HF component much broader than its LF counterpart. In the opposite case of relatively strong dispersion ($\gamma \geq \beta$), the input pulse tends to generate a single soliton with approximately equal widths of its components. While the solitons considered in subsection 3.1 are dominated by the dynamics of the LF component, and, as the usual KdV solitons, run only in one (positive) direction, which corresponds to *supersonic solitons* in the Davydov's model [20], simulations presented in subsection 3.2 demonstrate that, in the same system, HF-dominated solitons travel with in the opposite (negative) direction, corresponding to *subsonic* Davydov's solitons [20]). Collisions between stable solitons are studied in Section 3.3, by means of direct simulations. It is concluded that the collisions are inelastic, making a larger soliton stronger, and attenuating a smaller one. Robust intrinsic oscillations are excited in the colliding solitons, suggesting the existing of internal modes in them.

Analytical results are reported in Section 4. First, in subsection 4.1 a new family of approximate analytical two-component soliton solutions, with two free parameters, is derived for an arbitrary relative strength of the nonlinearity ($\beta$) and dispersion ($\gamma$) of the LF waves, assuming weak feedback of the HF component onto the LF one. In agreement with the numerical findings, the approximate analytical solutions have the width of the HF component which is much larger than that of the LF in the case of weak dispersion ($\gamma << \beta$). In subsection 4.2 a one-parameter (non-generic) family of exact two-component bright solitons with squared-sech components is produced, under condition of relatively weak dispersion, $\gamma < 3\beta$. Perturbed dynamics of the solitons, induced by a spatial shift of their HF component with respect to the LF one (this is a critical perturbation which may cause instability of the solitons) and collisions between the analytically found solitons are also considered in Section 4 by means of direct simulations.

While the two-component solitons considered in subsection 4.1 and 4.2 are LF-dominated ones, which travel with positive velocities, in subsection 4.3 we develop a quasi-adiabatic approximation, which makes it possible to predict HF-dominated solitons, which run with negative velocities. As mentioned above, the positive and negative velocities correspond to the supersonic and subsonic DS solitons, in terms of the Davydov's model [20]. The paper is concluded by Section 5.

## 2. The system of coupled LS-KdV equations

We start by considering the unidirectional copropagation of the complex HF wave field, $U(\xi,t)\exp(ik_0\xi - i\omega_0 t)$, with a slowly varying envelope, $U(x,t)$, and the real LF field, $n(x,t)$ (effectively, it may be considered as a local perturbation of the refractive index acting on the HF component), in the framework of the LS and KdV equations, coupled by the usual (quadratic) terms [9,15,16]:

$$2i\frac{\partial U}{\partial t} - \frac{\partial^2 U}{\partial \xi^2} + 2nU = 0, \tag{1}$$

$$2\frac{\partial n}{\partial t} - 6\beta\frac{\partial(n^2)}{\partial \xi} + \gamma\frac{\partial^3 n}{\partial \xi^3} = -\varepsilon\frac{\partial(|U|^2)}{\partial \xi}, \tag{2}$$

where $t$ and $\xi$ are the temporal and spatial variables, $\beta > 0$ and $\gamma > 0$ are the above-mentioned nonlinearity and dispersion coefficients of the LF waves, and $\varepsilon$ is the HF-induced striction (ponderomotive) coefficient.

Note that scaling transformation $\xi = \gamma\tilde{\xi}, t = \gamma^2\tilde{t}, U = \gamma^{-2}\varepsilon^{-1/2}\tilde{U}, n = \gamma^{-2}\tilde{n}$ transforms Eqs. (1) and (2) into the system with $\varepsilon = \gamma = 1$ and $\beta$ replaced by $\beta/\gamma$, i.e., the system with the single free constant, exactly the one (ratio $\beta/\gamma$) which plays the role of the control parameter in the present analysis. Nevertheless, for producing particular results below, we prefer to keep the system in the original form of Eqs. (1) and (2), as it will be convenient to vary all the three parameters, $\beta$, $\gamma$, and $\beta/\gamma\varepsilon$.

The system can be represented in the Lagrangian form, if the LF field is defined in terms of its potential, $n \equiv \partial v / \partial \xi$ (cf. Ref. [24], which employed a similar definition, $n \equiv \partial^2 v / \partial \xi^2$, to derive the Lagrangian representation for the bidirectional Zakharov's system, with even-order derivatives in the LF equation), the Lagrangian being

$$L = \int_{-\infty}^{+\infty} \left[ i\left(U \frac{\partial U^*}{\partial t} - U^* \frac{\partial U}{\partial t}\right) - \left|\frac{\partial U}{\partial \xi}\right|^2 - 2 \frac{\partial v}{\partial \xi} |U|^2 \right. \\ \left. - \frac{2}{\varepsilon} \frac{\partial v}{\partial \xi} \frac{\partial v}{\partial t} + \frac{2\beta}{\varepsilon}\left(\frac{\partial v}{\partial \xi}\right)^3 + \frac{\lambda}{\varepsilon}\left(\frac{\partial^2 v}{\partial \xi^2}\right)^2 \right] d\xi,$$

where the asterisk stands for the complex conjugate field. Further, the application of the Legendre transformation to the Lagrangian produces the respective Hamiltonian (written in terms of $n$, rather than $v$), which is a dynamical invariant of the underlying system of Eqs. (1) and (2):

$$H = \int_{-\infty}^{+\infty} \left[ \left|\frac{\partial U}{\partial \xi}\right|^2 + 2\frac{\partial v}{\partial \xi}|U|^2 - \frac{2\beta}{\varepsilon}n^3 - \frac{\gamma}{\varepsilon}\left(\frac{\partial n}{\partial \xi}\right)^2 \right] d\xi.$$

Other dynamical invariants are the total momentum, which is generated from the Lagrangian by the Noether theorem,

$$P = \int_{-\infty}^{+\infty} \left[ i\left(U \frac{\partial U^*}{\partial \xi} - U^* \frac{\partial U}{\partial \xi}\right) - \frac{2}{\varepsilon}n^2 \right] d\xi$$

as well as the wave action of the HF component and the total mass of the LF one:

$$N = \int_{-\infty}^{+\infty} |U(\xi)|^2 d\xi, \quad M = \int_{-\infty}^{+\infty} n(\xi) d\xi.$$

With zero HF component, $U = 0$, the system of Eqs. (1) and (2) is reduced to the KdV equation with the respective soliton solution:

$$n = -\frac{\gamma}{\beta\Delta^2} \operatorname{sech}^2\left(\frac{\xi - Vt}{\Delta}\right), \quad V = \frac{2\gamma}{\Delta^2}, \tag{3}$$

where $\Delta$ is an arbitrary width, and $V$ is the respective velocity.

When both the HF and LF components are present, steadily moving soliton solutions can be looked for in the following form:

$$U(t,\xi) = \psi(\eta)\exp(-iV\xi - i\Omega t), n = n(\eta), \eta \equiv \xi - Vt, \tag{4}$$

where real $\psi$ and $n$ are determined by the ordinary differential equations,

$$\frac{d^2\psi}{d\eta^2} - 2n\psi - \lambda\psi = 0, \quad \lambda = 2\Omega + V^2, \tag{5}$$

$$\gamma \frac{d^2 n}{d\eta^2} - 6\beta n^2 - 2Vn + \varepsilon\psi^2 = 0, \tag{6}$$

Soliton solutions should be exponentially localized. As it follows from Eqs. (5) and (6), this condition implies $V > 0$ (i.e., the KdV solitons run in the single (positive) direction) and $\lambda > 0$.

It is commonly known that the Schrödinger and KdV equations are invariant with respect to the Galilean transformations (GTs), which take different forms for these equations. In the Schrödinger equation, moving wave forms are Galilean transforms of quiescent ones, while KdV solitons (3) exist only with nonzero velocity. In the application to coupled Eqs. (1) and (2), the GT transforms a given soliton, with velocity $V_1 > 0$, into one with any other velocity, $V_2 > 0$:

$$\psi_2(\eta) = (V_2/V_1)\psi_1\left(\sqrt{V_1/V_2}\,\eta\right), \quad n_2(\eta) = (V_2/V_1)n_1\left(\sqrt{V_1/V_2}\,\eta\right),$$

$$\Omega_2 = \frac{V_2}{V_1}\Omega_1 + \frac{1}{2}V_2(V_1 - V_2).$$

The following explicit (exact and approximate) soliton solutions of Eqs. (1) and (2) were obtained in the previous works, as mentioned in the Introduction:

(i) For $\gamma = \beta$, the two-component DS soliton solution, found in the context of the original Davydov's model for long polymer molecules [18] and in other contexts [19]:

$$U = A\,\text{sech}\left[\sqrt{\lambda}(\xi - Vt)\right]\exp(-i\lambda t - iV\xi), \quad n = -\lambda\,\text{sech}^2\left[\sqrt{\lambda}(\xi - Vt)\right], \tag{7a}$$

where

$$V \equiv 2\beta\lambda - \varepsilon A^2/(2\lambda), \tag{7b}$$

with two arbitrary real parameters $\lambda$ and $A$. This particular exact solution is invariant with respect to the above Galilean transformation. Note that, unlike the KdV soliton (3), whose velocity cannot change its sign, the velocity of the exact DS solution (7a), as given by Eq. (7b), may be both positive and negative, provided that the product of the nonlinearity coefficients is positive, $\beta\varepsilon > 0$.

(ii) For the same case, $\beta = \gamma$, solitons supported by finite background were found [20]. They include modes with the dark-soliton or "bubble" [24] structure in the HF component, and a "bubble" in the LF wave, as well as new two-hump dark solitons.

(iii) For $\gamma \neq \beta$ and a small amplitude of the HF field, $|U|/n \ll 1$, approximate multihump solitons were found in Ref. [10], in terms of a system originating from a lattice model.

(iv) Neglecting the feedback action of the HF components on the LF waves, which induce a potential well in the HF equation, localized linear stationary pulses of the HF field may be bound states of the usual linear Schrödinger equation [16].

## 3. Numerical results

### 3.1. The evolution of input pulses dominated by the LF component

To gain insight into the behavior of the coupled Schrödinger-KdV system with different relative strengths of the dispersion and nonlinearity in the KdV component, we start with systematic simulations of Eqs. (1) and (2). For this purpose, the finite-difference scheme was used, implemented by means of the Maple software shell.

Here, we display results which adequately represent the typical situation, by fixing $\varepsilon = 1/10$, $\beta = 1/5$, and varying $\gamma$. The input was taken in the form of a generic two-component pulse, which follows the structure of the DS soliton (4):

$$U(\xi, 0) = \text{sech}\,\xi, \quad n(\xi, 0) = -\text{sech}^2\xi. \tag{8}$$

At $\gamma < \beta$, this input tends to split into several two-component solitons (Figs. 1 and 3), with the width of the HF component larger than that of the LF field (see Figs. 2 and 4). These results may be compared to those reported in Ref. [10], where formation of complexes built of several solitons, but with zero velocities, was considered in the framework of a system in which the LF component was governed by an equation of the Boussinesq type. Unlike the present situation, those complexes might form stationary multi-soliton bound states.

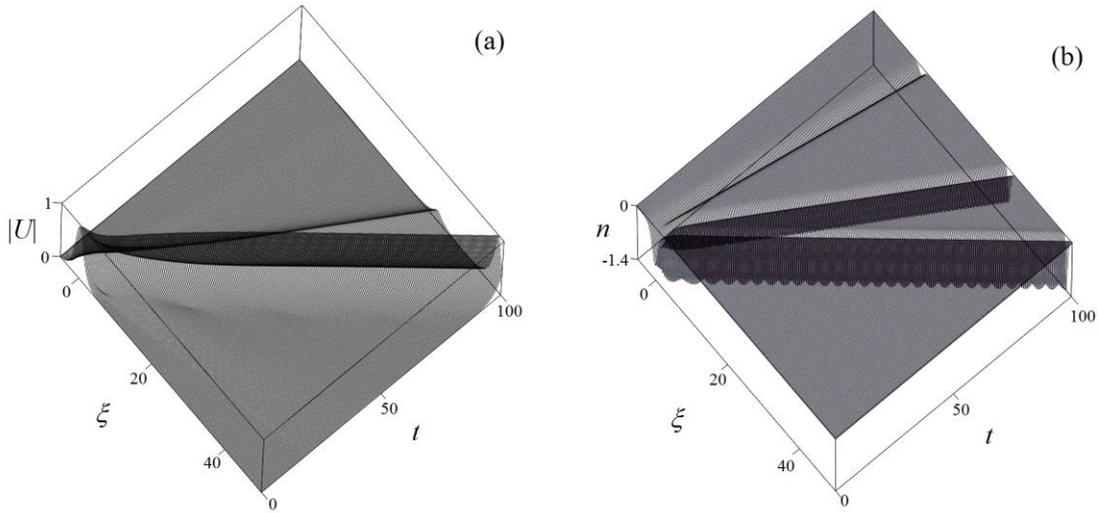

Fig. 1. Results of simulations of the system (1)-(2) with $\gamma = 1/30$ (i.e., $\gamma/\beta = 1/6$) for the spatio-temporal distributions of the HF field, $|U(\xi,t)|$ (a), and LF field, $n(\xi,t)$ (b), produced by initial pulse (8). Other parameters are $\varepsilon = 1/10$ and $\beta = 1/5$. The evolution generates three solitons, as can be clearly seen in Fig. 2.

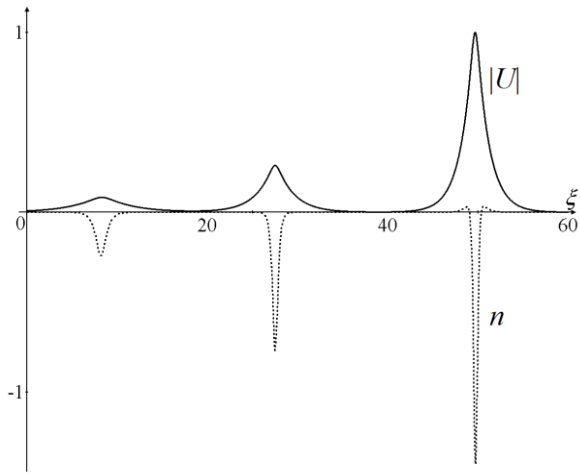

Fig. 2. Results of the simulation from Fig. 1, i.e., with $\gamma/\beta = 1/6$, at $t = 100$. The solid and dotted curves show, respectively, the HF and LF fields, $|U(\xi)|$ and $n(\xi)$.

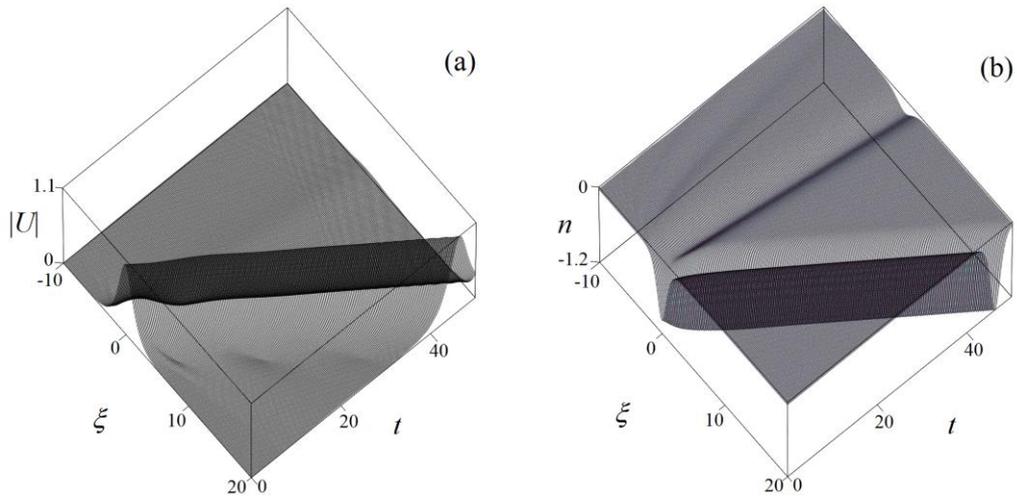

Fig. 3. The same as in Fig. 1, but for $\gamma = 1/10$, i.e., $\gamma/\beta = 1/2$. In this case, two solitons are produced by the evolution, as is additionally demonstrated by Fig. 4.

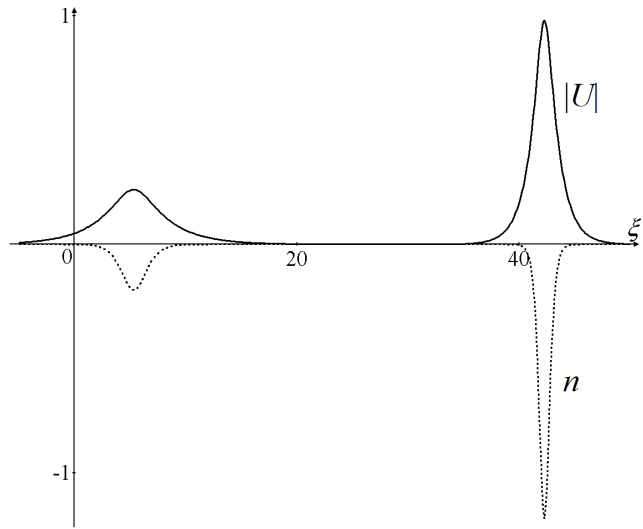

Fig. 4. The same as in Fig. 2, but for $\gamma = 1/10$, i.e., $\gamma/\beta = 1/2$.

At $\gamma \geq \beta$, the results of the evolution of input (8) are essentially different, generating a single two-component soliton (Fig. 5), with approximately equal widths of both components (see Fig. 6). In all the cases, the emerging solitons are, obviously, stable ones (otherwise, they would not self-trap in the direct simulations).

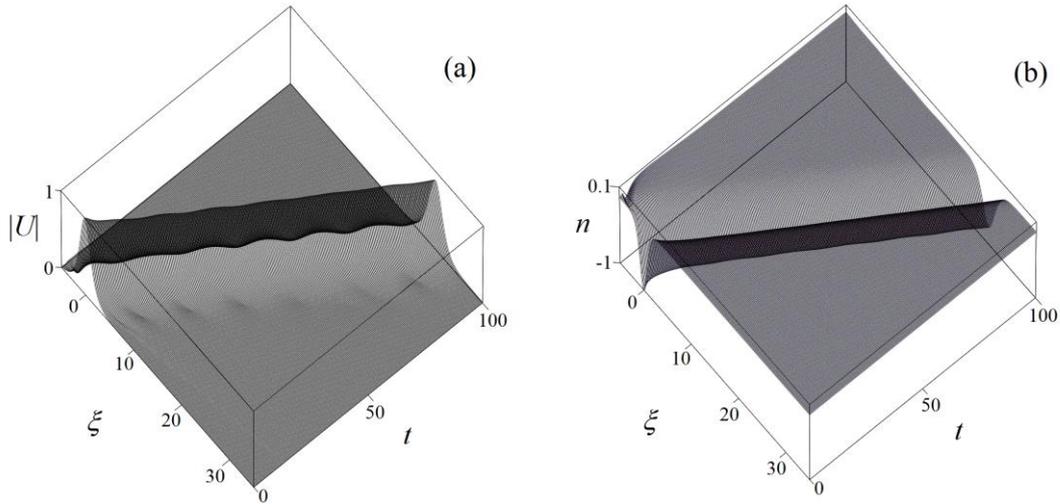

Fig. 5. The same as in Fig. 1, but for $\gamma = 2/5$, i.e., $\gamma/\beta = 2$.

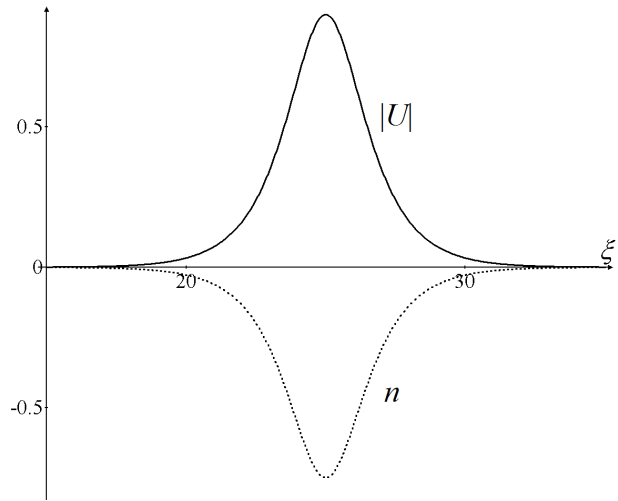

Fig. 6. The same as in Fig. 2, but for $\gamma = 2/5$.

To corroborate the generic character of the conclusions formulated above, we additionally display results of the evolution of the input with a larger amplitude of the LF component, while keeping the same values $\varepsilon = 1/10$, $\beta = 1/5$ as above:

$$U(\xi,0) = \mathrm{sech}\,\xi, \quad n(\xi,0) = -(3/2)\mathrm{sech}^2\xi, \qquad (9)$$

cf. Eq. (8). Outputs of the simulations are displayed in Fig. 7 for $\gamma = 1/30$ (a) and $\gamma = 2/5$ (b). They clearly corroborate the above conclusions: on the one hand, the formation of multiple solitons with the HF component much broader than its LF counterpart in the case of $\gamma/\beta \ll 1$ (Fig. 7(a)), and, on the other hand, the creation of the single soliton with approximately equal widths of its components in the case of $\gamma/\beta \geq 1$, see Fig. 7(b).

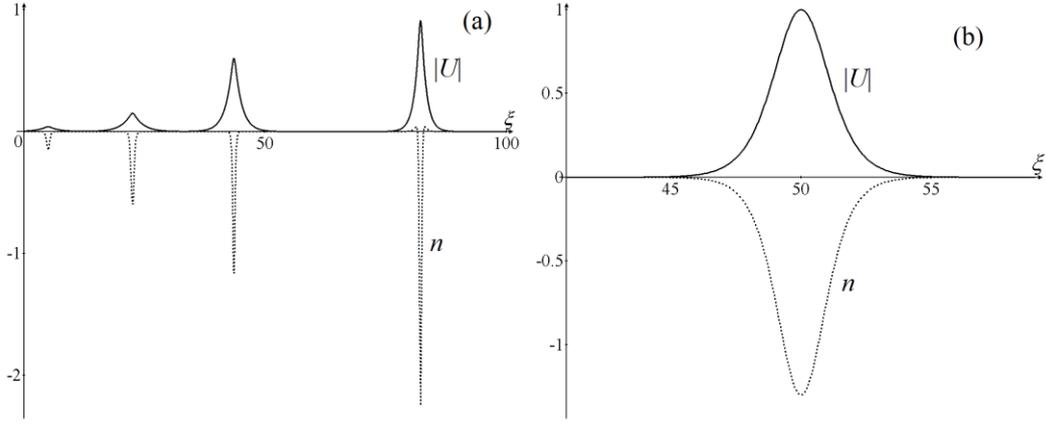

Fig. 7. Results of simulations of the system (1)-(2) with input (9) and parameters (a) $\gamma = 1/30$, i.e., $\gamma/\beta = 1/6$ (cf. Fig. 2), and (b) $\gamma = 2/5$, i.e., $\gamma/\beta = 2$ (cf. Fig. 6), at $t = 100$. Other parameters are $\varepsilon = 1/10$ and $\beta = 1/5$.

## 3.2. The evolution of input pulses dominated by the HF component

Very different results are produced by the input which includes solely the HF component,
$$U(\xi,0) = U_0 \text{sech}\xi, \quad n(\xi,0) = 0. \qquad (10)$$
For the same parameters as used above, i.e., $\varepsilon = 1/10$ and $\beta = 1/5$, the output of the simulations is displayed in Figs. 8 and 9 (with $U_0 = 2$ in Eq. (10) and $\gamma = 1/10$), which demonstrate that the HF component quickly develops a soliton structure, while the LF field features a quasi-localized undulatory shape, with a long dispersive tail attached to it on the right-hand side. A crucial difference of this result from those displayed above in Figs. 1, 3, and 5 is that, in the present case, the two-component quasi-soliton moves with a *negative velocity* (like *subsonic solitons*, in terms of the Davydov's model [20]). Below (in subsection 4.3), we demonstrate that the LF field in this configuration may be understood as a superposition of a soliton's LF component, obtained in the quasi-adiabatic approximation [25], and a dispersive quasi-linear wave. A characteristic feature of the quasi-adiabatic solitons is that their velocity is negative, as indeed observed here.

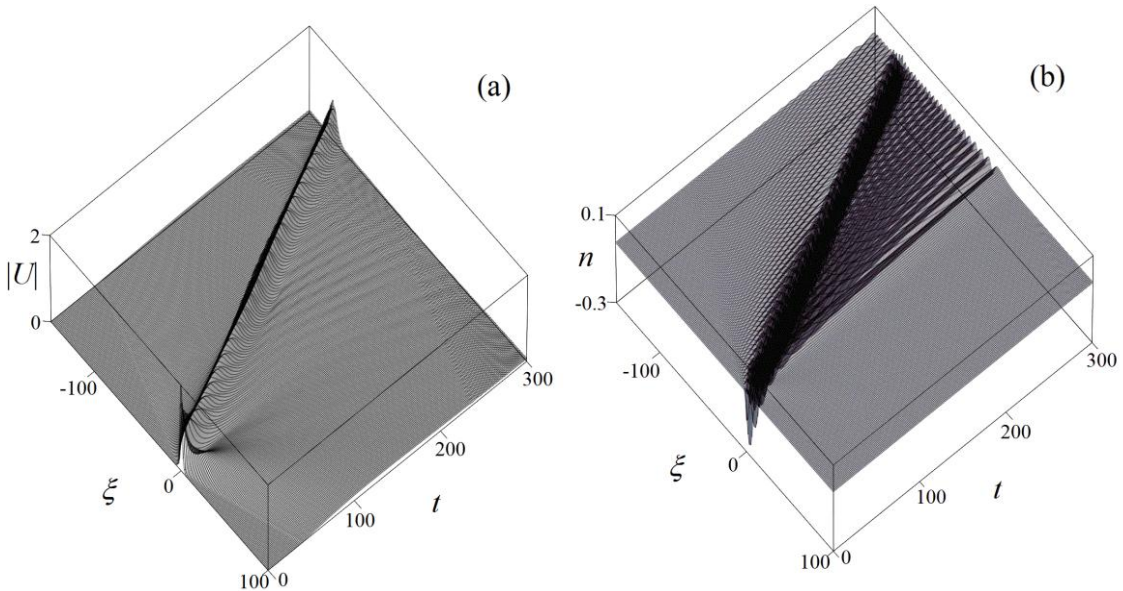

Fig. 8. Results of simulations of the system of Eqs. (1)-(2), produced by input (10) with $U_0 = 2$ and zero LF component. Other parameters are the same as in the case shown in Figs. 3 and 4, i.e., $\gamma = 1/10$, $\varepsilon = 1/10$, and $\beta = 1/5$.

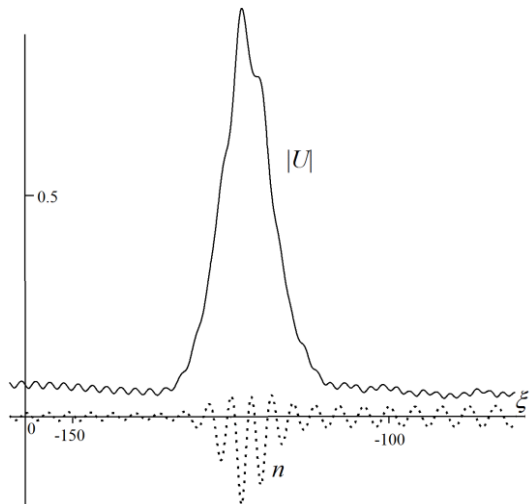

Fig. 9. The outcome of the simulations, shown in Fig. 8, at $t = 300$. The solid and dotted curves show, respectively, the HF and LF fields, $|U(\xi)|$ and $n(\xi)$.

## 3.3. Soliton-soliton collisions

Interactions between stable solitons are an issue of obvious interest to physical realizations of the system. Simulations of soliton-soliton collisions also help to understand general dynamical properties of the system under the consideration.

Here we consider collisions between different solitons, with components $U_1(\xi), n_1(\xi)$ and $U_2(\xi), n_2(\xi)$, produced as outlined in the previous subsection. Actually, these two solitons were taken as ones generated from inputs (8) and (9), respectively. They were initially placed at a sufficiently large distance, $\xi_0 = 20$, between their centers. We display results of simulations of the collisions for the same parameters as considered above, i.e., $\varepsilon = 1/10$, $\beta = 1/5$ and different values of $\gamma$.

For $\gamma = 1/30$ (i.e., $\gamma/\beta = 1/6$), the first and second colliding solitons are the largest ones from Figs. 2 and 7(a), respectively, with velocities $V_1 = 0.5$ and $V_2 = 0.8$. The simulation, displayed in Fig. 10, demonstrates that the collision is inelastic: although the solitons pass through each other, the initially larger one becomes still larger, while the smaller one becomes still smaller. The collision also causes excitation of conspicuous intrinsic vibrations in the larger soliton, which suggests the existence of an intrinsic mode in the solitons.

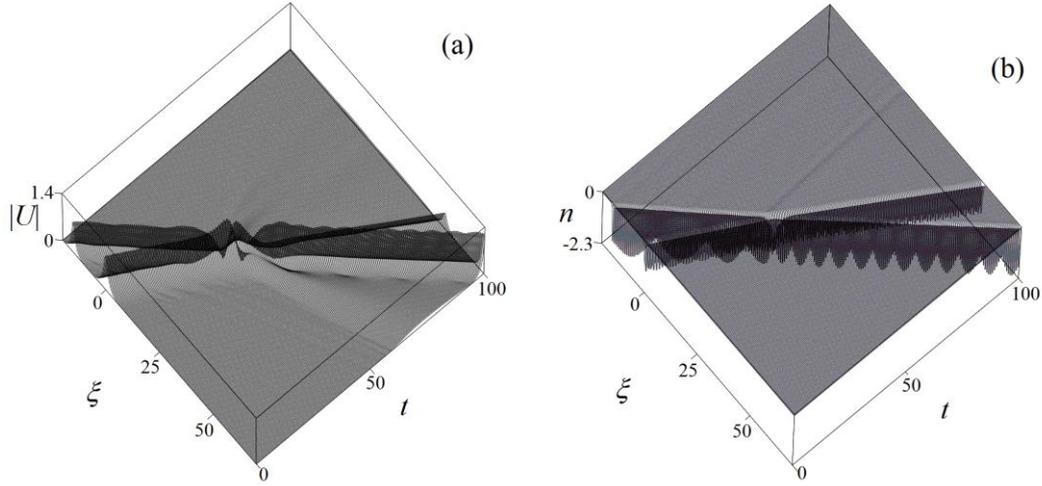

Fig. 10. Results of simulations of Eqs. (1) and (2) with $\gamma = 1/30$ ($\gamma/\beta = 1/6$) for the collision between the largest solitons from Figs. 2 and 7(a), moving with velocities $V_1 = 0.5$ and $V_2 = 0.8$, respectively.

For $\gamma = 2/5$ (i.e., $\gamma/\beta = 2$), the collision of solitons from Figs. 4 and 7(b), moving with velocities $V_1 = 0.25$ and $V_2 = 0.5$, respectively, is displayed in Fig. 9. The collision demonstrates inelasticity similar to that observed in Fig. 8.

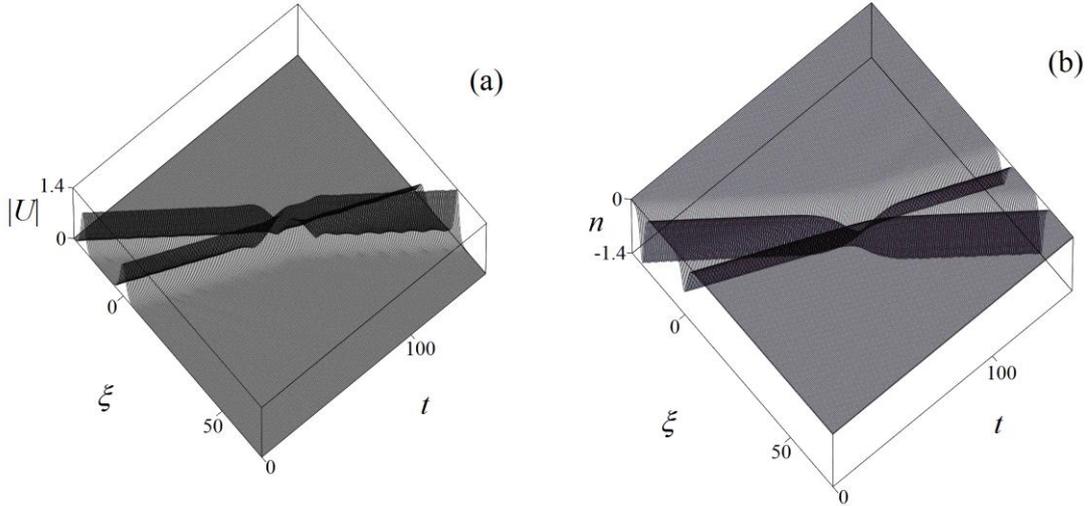

Fig. 11. The same as in Fig. 10, but for $\gamma = 2/5$ ($\gamma/\beta = 2$). In this case, the colliding solitons are ones from Figs. 4 and 7(b), moving with velocities $V_1 = 0.25$ and $V_2 = 0.5$, respectively.

## 4. Analytical results

### 4.1 Approximate soliton solutions with the dominant LF component

Most essential results reported in this work are analytical ones. First, in the particular case when the HF field is absent, i.e., $U = 0$, an obvious solution of Eq. (6) is

$$n = -\frac{\gamma}{\beta \Delta^2} \operatorname{sech}^2\left(\frac{\eta}{\Delta}\right), \quad \Delta \equiv \sqrt{\frac{2\gamma}{V}}. \tag{11}$$

Substituting this expression for the LF field in Eq. (5), and neglecting the ponderomotive feedback of the HF field on the LF waves, i.e., setting $\varepsilon = 0$ in Eqs. (2) and (6), we arrive at the stationary Schrödinger equation with the Pöschl-Teller potential ($\mathrm{sech}^2\rho$) (cf. a similar approach developed earlier in work [23]):

$$\frac{d^2\psi}{d\rho^2} - k^2\psi + \frac{2\gamma}{\beta \cosh^2\rho}\psi = 0, \qquad (12a)$$

$$\rho \equiv \eta/\Delta, \quad k^2 \equiv \lambda\Delta^2. \qquad (12b)$$

The ground-state eigenvalue of Eq. (12a) is

$$k = \frac{1}{2}\left(\sqrt{1+8\frac{\gamma}{\beta}} - 1\right), \qquad (13)$$

with the corresponding *exact* wave function [26],

$$\psi(\rho) = A\,\mathrm{sech}^k\rho, \qquad (14)$$

where $A$ is a free parameter. In the limit cases of $\gamma \ll \beta$ and $\gamma \gg \beta$ (weak or strong dispersion of the KdV equation in comparison with the nonlinearity), the eigenvalue becomes, respectively, $k \approx 2\gamma/\beta$, and $k \approx \sqrt{2\gamma/\beta}$.

The analytical wave function produced by Eq. (14) is plotted in Fig.12 (solid curves) for $A = 1$ and different values of the dispersion/nonlinearity ratio: $\gamma/\beta = 1/6$, $\gamma/\beta = 2$, and $\gamma/\beta = 8$ As expected, for $\gamma/\beta \ll 1$ the HF component is much broader than the LF one, which supports it by means of the effective potential in Eq. (10), while in the opposite limit, $\gamma/\beta \gg 1$, the HF component becomes narrower than the LF counterpart.

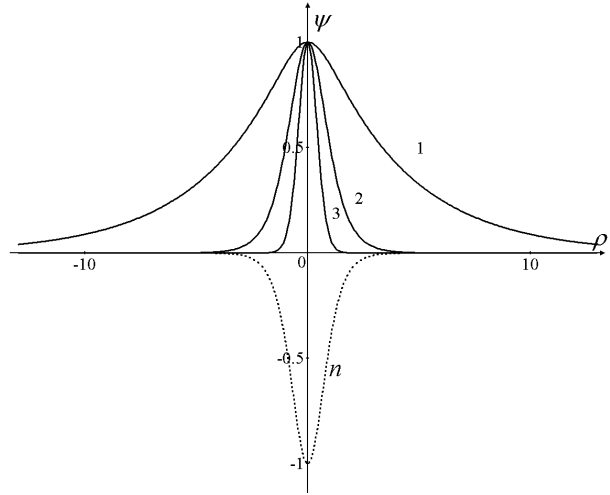

Fig.12. Analytically obtained wave functions (14) (solid curves) for $A = 1$ and different values of the dispersion/nonlinearity ratio: curve 1 corresponds to value $\gamma/\beta = 1/6$ ($k \approx 0.26$); 2 - $\gamma/\beta = 2$ ($k \approx 1.56$); 3 - $\gamma/\beta = 8$ ($k \approx 5.56$). The dotted bottom curve shows the LF profile from Eq. (10), $n = -\mathrm{sech}^2\rho$.

In Fig. 13, the analytical waveform (14), shown by dashed curves, is compared to the HF component (solid curves) of the numerically found solutions for $|U(\eta)|$: (a) with $\gamma = 1/30$, corresponding to the largest soliton generated from input (8) (see Fig. 2); (b) with $\gamma = 2/5$, generated from the same initial pulse (Fig. 6). In Fig. 13(a), Eqs. (14) and (15) yield $k \approx 0.26$, $\rho \equiv \eta / \Delta \approx \eta / 0.3$, and $A = 1$ provides the best fit of the analytical result to the numerical counterpart. In Fig. 13(b), $k \approx 1.56$, $\rho \equiv \eta / \Delta \approx \eta / 0.5$, and the best fit is provided by $A = 0.9$. Thus, good agreement between the analytical and numerical results in observed in the figure, if the fitting parameter $A$ is chosen appropriately.

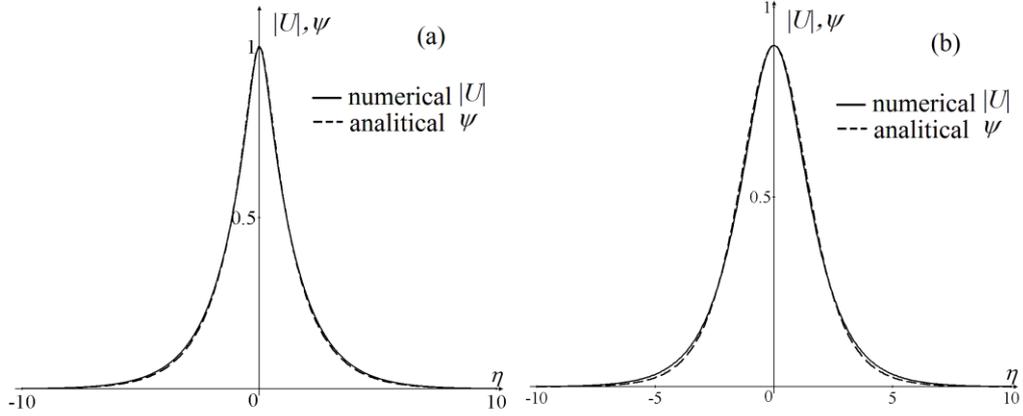

Fig.13. Solid curves depict components $|U(\eta)|$ of the numerically found solution: (a) for $\gamma = 1/30$, taken as the tallest soliton in Fig. 2 (a); (b) for $\gamma = 2/5$, taken as the soliton from Fig. 6. Dashed curves: the respective analytical wave functions $\psi$ from Eq. (14), with appropriately adjusted free parameter $A$ (as explained in the main text, $k \approx 0.26$ in (a), and $k \approx 1.56$ in (b)). Other parameters are $\varepsilon = 1/10$ and $\beta = 1/5$.

The feedback of HF mode (14) on the LF component, which was neglected in the above analysis, can be taken into account perturbatively, adding the respective correction, $n_1$, to solution (10):

$$n = -\frac{\gamma}{\beta \Delta^2} \text{sech}^2\left(\frac{\eta}{\Delta}\right) + n_1. \qquad (15)$$

Substituting this in Eq. (6), neglecting the term $\sim n_1^2$, and returning to variable $\rho = \eta / \Delta$ as per Eq. (12), we arrive at a linear inhomogeneous equation:

$$\frac{d^2 n_1}{d\rho^2} + \frac{12 n_1}{\cosh^2 \rho} - 4n_1 = -\frac{\varepsilon \Delta^2 A^2}{\gamma \cosh^{2k} \rho}. \qquad (16)$$

For $k \ll 1$, which corresponds to $\gamma / \beta \ll 1$, pursuant to Eq. (13), Eq. (16) produces a broad correction to the narrow LF mode. In the outer zone of the solution, $\rho > 1$, where the correction is a dominant term, it can be easily obtained from Eq. (16), in which one may neglect the first two terms on the left-hand side:

$$n_1 \approx \frac{\varepsilon \Delta^2 A^2}{4\gamma} \text{sech}^{2k}\left(\frac{\eta}{\Delta}\right). \qquad (17)$$

The feedback-corrected LF mode (15), with the correction taken according to Eq. (17), is compared to its numerically found counterpart in Fig. 14, for the abovementioned values of the parameters: $\gamma = 1/30$ ($k \approx 0.26$), $\Delta \approx 0.3$ and $A = 1$ (dashed curve). In this case too, the agreement of the analytical prediction with the numerical solution is good. Note that correction (17) is essential, as, without it, the zeroth-order analytical approximation does not produce local maxima and positive tails of the LF mode.

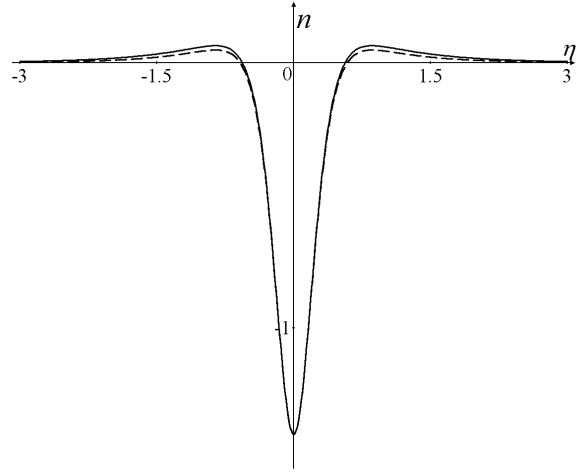

Fig.14. The solid curve depicts the LF component of the numerically found soliton with $\gamma = 1/30$, corresponding to the tallest one in Fig. 2 (initially generated by input (8)). The dashed curve represents the respective analytical approximation given by Eqs. (15) and (17), for $\gamma = 1/30$ ($k \approx 0.26$), $\Delta \approx 0.3$ and the fitting parameter $A = 1$. Other parameters are $\varepsilon = 1/10$ and $\beta = 1/5$.

## 4.2. Exact soliton solutions and numerical tests of their stability and collisions

The feedback of the HF field on the LF component can be taken into account in an *exact form* in the particular case of $k = 2$ in Eq. (14), which corresponds to

$$\gamma / \beta = 3 \tag{18}$$

in terms of Eq. (13). In this case, we replace Eq. (10) by

$$n = -N \operatorname{sech}^2(\eta/\Delta), \tag{19}$$

with some amplitude $N$, that should be found, and the substitution of $\psi = A \operatorname{sech}^2 \rho$ back in Eq. (6) tells us that $\beta$ must be replaced by $\tilde{\beta} \equiv \beta - \dfrac{\varepsilon}{6} \dfrac{A^2}{N^2}$. Then, Eq. (18) must hold, with $\beta$ replaced by $\tilde{\beta}$, which yields

$$\frac{\varepsilon}{6} \frac{A^2}{N^2} = \beta - \frac{\gamma}{3}. \tag{20}$$

Further, it follows from the comparison of Eqs. (10) and (19) that

$$N = \frac{\gamma}{\tilde{\beta} \Delta^2} \equiv \frac{\gamma}{\left(\beta - \dfrac{\varepsilon}{6} \dfrac{A^2}{N^2}\right) \Delta^2}. \tag{21}$$

Finally, combining Eqs. (20) and (21), one can find the amplitudes of the HF and LF and fields, as a part of the exact solution:

$$N = \frac{3}{\Delta^2}, A^2 = \frac{18(3\beta - \gamma)}{\varepsilon \Delta^4}. \tag{22}$$

Thus, the final form of the family of the exact solutions to Eqs. (5) and (6) is

$$\psi = \frac{1}{\Delta^2} \sqrt{\frac{18(3\beta - \gamma)}{\varepsilon}} \operatorname{sech}^2\left(\frac{\eta}{\Delta}\right), \quad n = -\frac{3}{\Delta^2} \operatorname{sech}^2\left(\frac{\eta}{\Delta}\right), \tag{23}$$

where $\Delta$ is a free parameter of the family, while the soliton's velocity and the frequency of its HL component (see Eqs. (4) and (5)) are

$$V = 2\gamma / \Delta^2, \Omega = (2/\Delta^2)(1 - \gamma^2/\Delta^2). \tag{24}$$

In particular, for $\gamma = 3\beta$ the HF component of soliton (23) vanishes, and the solution reduces to the KdV soliton (3). The presence of the single free parameter in the family, $\Delta$, implies that it is not a generic soliton family, as the latter one should have two free independent parameters, representing the velocity and amplitude of solitons [21].

In fact, the exact solution given by Eqs. (23) and (24) may be considered as a bright version of a similar soliton reported in a very recent work [20], under the name of the "Davydov soliton of the second kind", alias a supersonic Davydov soliton, with a difference that the solution presented in Ref. [20] includes a nonzero background in the LF component.

Stability of exact solution (23) was tested by means of direct simulations of the underlying system of Eqs. (1) and (2). As a specific perturbation, the simulations included an initial shift, $\xi_0$, of the centers of the HF and LF components, i.e., the respective initial condition was taken as

$$U(\xi,0) = \frac{1}{\Delta^2}\sqrt{\frac{18(3\beta-\gamma)}{\varepsilon}}\operatorname{sech}^2\left(\frac{\xi-\xi_0}{\Delta}\right)\exp\left(-\frac{2i\gamma\xi}{\Delta^2}\right), \quad n(\xi,0) = -\frac{3}{\Delta^2}\operatorname{sech}^2\left(\frac{\xi}{\Delta}\right). \quad (25)$$

First, Fig. 15 demonstrates that, with $\xi_0 = 0$, the soliton remains completely stable in the course of the long evolution (in the present case, $t = 200$ corresponds, roughly, to ~ 40 HF diffraction times).

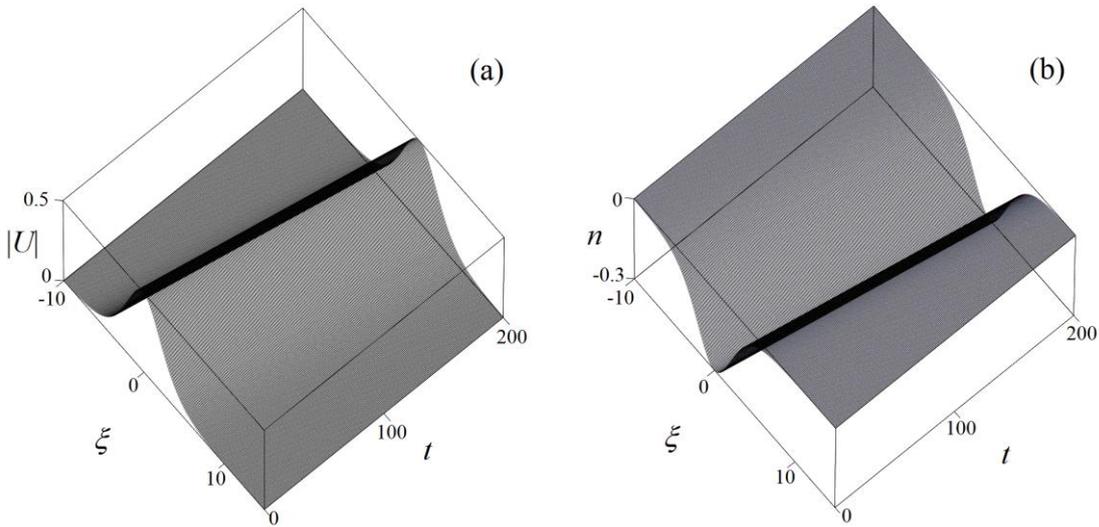

Fig.15. Results of the simulations of the system of Eqs. (1) and (2) with input (25) in the case of $\xi_0 = 0$ and $\Delta = 3$, which corresponds to the exact soliton given by Eqs. (23) and (24). Parameters are $\gamma = 1/10$, $\beta = 1/15$ and $\varepsilon = 1/10$. Panels (a) and (b) display the evolution of the HF and LF components, respectively.

Further, Fig. 16 displays results produced by the simulations of the evolution of input (25) with initial shift $\xi_0 = 0.5$ between the components. In this case, the disturbed soliton features robust long-period intrinsic vibrations, which suggests that the soliton is a truly stable bound state, and, plausibly, it supports a stable mode of internal excitations.

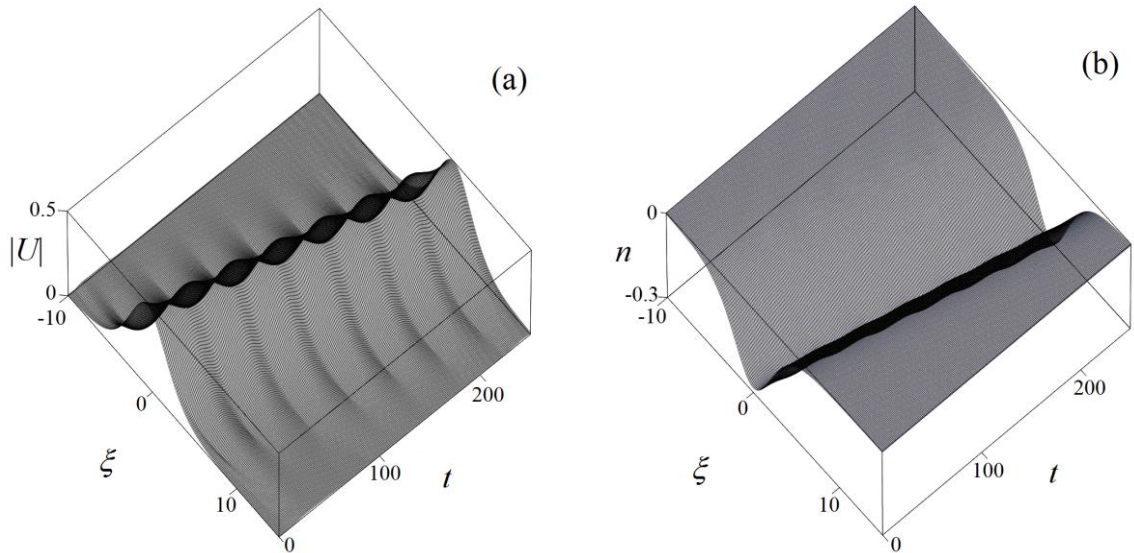

Fig. 16. The same as in Fig. 15, but for input (25) with initial shift $\xi_0 = 1/2$ between the HF and LF components.

As well as in the case of generic numerically found soliton solutions considered in the previous section, collisions between exact solitons are an issue of obvious interest to physical realizations of the system. We have performed systematic simulations of collisions between exact solitons with two different values of parameter $\Delta$, hence moving with different velocities (see Eq. (24)), initially separated by a sufficiently large distance.

Typical examples of the collisions are displayed in Figs. 17 and 18, with a sufficiently large initial distance between the solitons, for parameters $\gamma = 1/10$, $\beta = 1/20$, and $\varepsilon = 1/10$. In both cases, the faster soliton is taken with $\Delta_1 = 1$, while the slower one has $\Delta_2 = 2$ in Fig. 17 and $\Delta_2 = 1.5$ in Fig. 18. The numerical results demonstrate that the collisions are inelastic, similarly to those displayed in Figs. 10 and 11: initially larger soliton and smaller solitons come out still larger and smaller, respectively, and conspicuous intrinsic oscillations are excited in the colliding solitons.

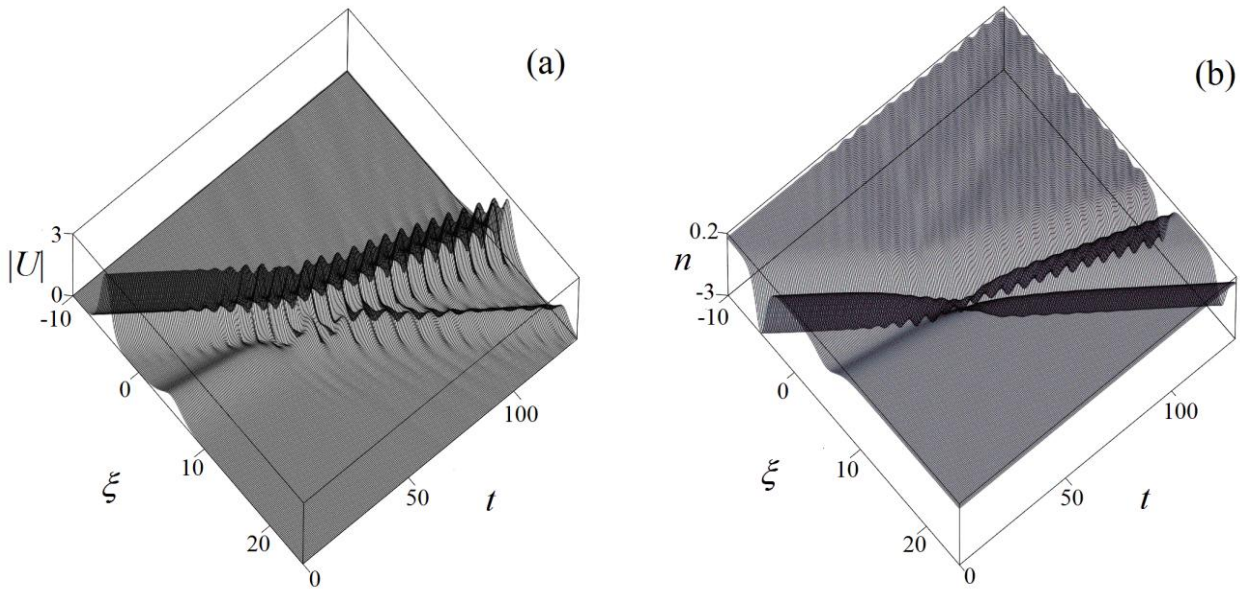

Fig.17. Collision of two exact solitons, given by Eqs. (23) and (24) with $\Delta_1 = 1$ and $\Delta_2 = 2$, at $\gamma = 1/10$, $\beta = 1/20$, $\varepsilon = 1/10$ and $\xi_0 = 5$. The initial separation between the solitons is 10. Panels (a) and (b) display the
evolution of the HF amd LF components, respectively.

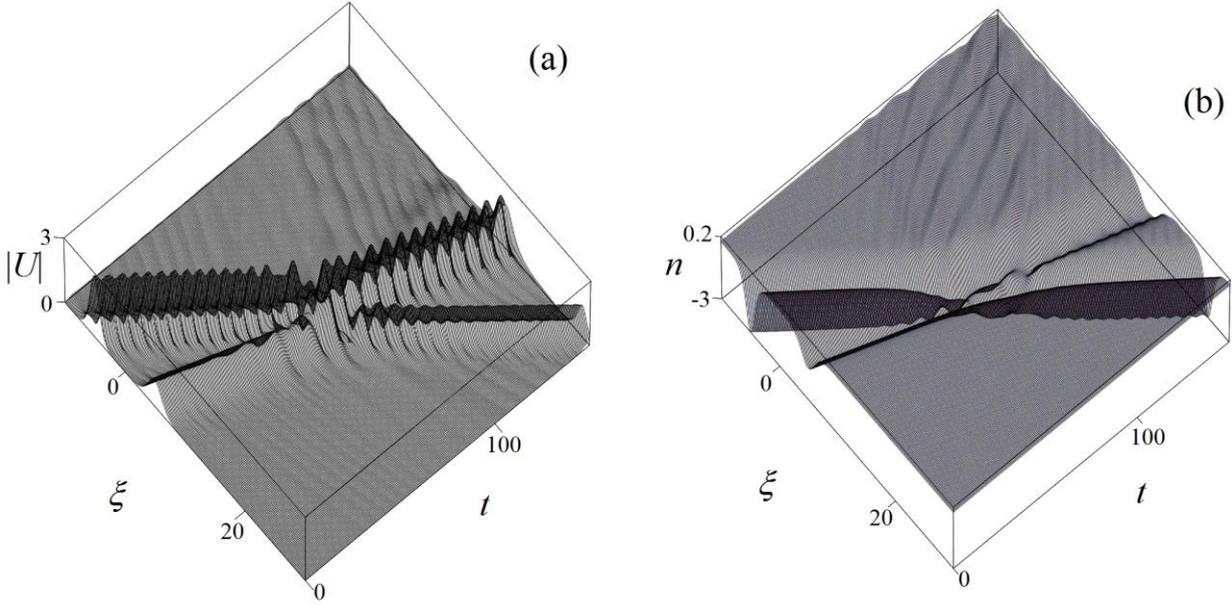

Fig.18. The same as in Fig. 17, but for $\Delta_2 = 3/2$.

**4.3. Approximate soliton solutions with the dominant HF component**

The adiabatic approximation is frequently applied to systems of the Zakharov's type [25]. In terms of Eq. (6), this approximation omits the second derivative and the term $\sim n^2$, thus reducing Eq. (6) to a simple relation,

$$n = (\varepsilon / 2V)\psi^2. \tag{26}$$

The substitution of this in Eq. (5) yields the stationary version of the cubic NLS equation,

$$\frac{d^2\psi}{d\eta^2} - \frac{\varepsilon}{V}\psi^3 - \lambda\psi = 0. \tag{27}$$

Obviously, Eq. (27) with $\lambda > 0$ gives rise to bright solitons,

$$\psi = \sqrt{-\frac{2V\lambda}{\varepsilon}}\,\text{sech}\left(\sqrt{\lambda}\eta\right), \tag{28}$$

but with the *negative velocity*, like in Fig. 8, and on the contrary to what is seen in Figs. 1, 3, 5, 10, 11, 15, 17, and 18. Thus, the solitons with the dominant HF component (the present derivation implies that the LF component (26) of the two-component soliton is small), unlike the complexes considered above, in which the LF component is the dominant one (or, at least, is comparable to its HF counterpart), move in the negative direction. In terms of the Davydov's model, solitons moving with negative velocities are called subsonic ones, on the contrary to the supersonic DSs, which run with positive velocities [20].

In fact, Fig. 9 suggests that the nonlinear term ($\sim n^2$) in Eqs. (2) and (6) for the LF field is indeed negligible, as the LF amplitude is very small, but the dispersive term may be essential in these equations. In that case, instead of adopting the simple adiabatic relation (26), one should solve the linearized version of Eq. (6),

$$\gamma\frac{d^2n}{d\eta^2} - 2Vn = -\varepsilon\psi^2. \tag{29}$$

It can be solved as the equation of motion for a driven harmonic oscillator, but the substitution of the solution in Eq. (5) leads to a cumbersome nonlinear integral equation. Instead, in the quasi-adiabatic approximation one may assume that the solution for $\psi$ and $n$ still has the form of

$$\psi = A\,\text{sech}\left(\sqrt{\lambda}\eta\right),\quad n = B\,\text{sech}^2\left(\sqrt{\lambda}\eta\right). \tag{30}$$

Then, substituting this ansatz in Eq. (29), one can approximately find $B$ from the integral version of the equation, multiplying it by $\text{sech}^2\left(\sqrt{\lambda}\eta\right)$ and performing the integration, $\int_{-\infty}^{+\infty} d\eta$. The result is

$$B = \frac{\varepsilon/2}{V+(2/5)\gamma\lambda}A^2. \tag{31}$$

Finally, the substitution of expressions (30) and (31) in Eq. (5) predicts a corrected expression for the amplitude of the HF components:

$$A = \sqrt{-\frac{2\lambda}{\varepsilon}\left(V+\frac{2}{5}\gamma\lambda\right)} \tag{32}$$

cf. Eq. (28). This results implies that the velocity of the HF-dominated two-component soliton is not merely negative, but its absolute value must be large enough: $V < -(2/5)\gamma\lambda$.

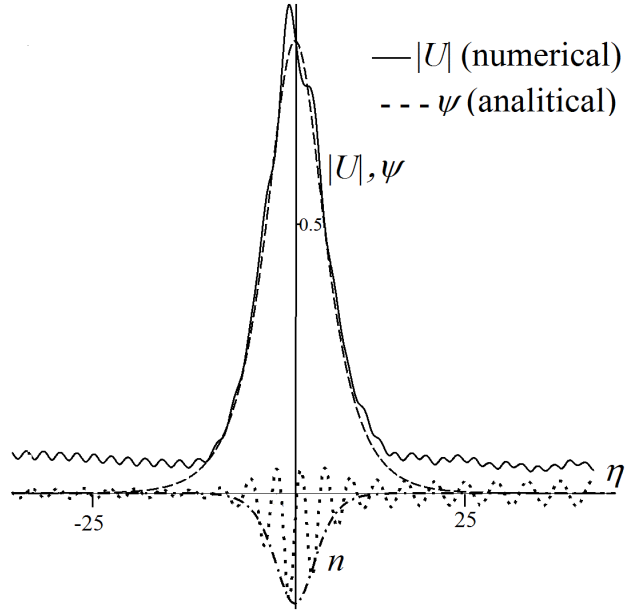

Fig. 19. Solid and dotted curves depict components $|U(\eta)|$ and $n(\eta)$ of the numerically found solution for $\gamma = 1/10$, taken from Fig. 9. Dashed and dashed-dotted curves: the respective analytical profiles of $\psi(\eta)$ and $n(\eta)$, as predicted by Eq. (30), (32) and (30), (31), respectively, with $\lambda = 1/12$ and $V = -2/5$. Other parameters are $\varepsilon = 1/10$ and $\beta = 1/5$.

Finally, as concerns the explanation of the numerical results displayed in Figs. 8 and 9, which correspond to initial conditions (10), the solution of the corresponding linearized (with respect to $n$) version of Eq. (2) may be looked for as

$$n(\xi,t) = n_{\text{sol}}(\eta) + n_{\text{rad}}(\xi,t), \tag{33}$$

where $n_{\text{sol}}(\eta)$ is the soliton's component approximated by Eqs. (30) and (31), and $n_{\text{rad}}(\xi,t)$ is the radiation (dispersion-wave) component, governed by the homogeneous linearized version of Eq. (2), taken without the term $\sim \varepsilon$. According to the initial condition imposed by Eq. (10), we have $n_{\text{rad}}(\xi) = -n_{\text{sol}}(\xi)$ at $t = 0$, and the undulating component of the LF field in Fig. 9 may be realized as the result of the evolution of $n_{\text{rad}}(\xi, t)$, governed by the linear dispersive homogenous equation. This conjecture is well corroborated by Fig. 19, which displays the comparison of the numerically generated fields from Fig. 9 with the analytical approximation based on Eqs. (30)-(32). It is seen that the HF pulse is accurately approximated by Eqs. (30) and (32) - naturally, except for the radiation tails. The envelope of the LF is also well approximated by Eqs. (30)-(32), i.e., by the first term in Eq. (33), while the intrinsic undulations under the envelope and small-amplitude tails attached to it correspond to the radiation component in Eq. (33).

## 5. Conclusion

The main objective of this work is to extend the study of solitons in coupled Schrödinger-KdV system to the case of an arbitrary relative strength of the nonlinearity and dispersion of the LF (low-frequency) component ($\beta$ and $\gamma$, respectively). New solitons have been reported in this system. For the case of weak dispersion, $\gamma \ll \beta$, the numerically simulated evolution of the generic input tends to create several two-component solitons, with the HF (high-frequency) component much broader than its LF counterpart. For strong dispersion, $\gamma \geq \beta$, the initial pulse typically gives rise to a single two-component soliton, with approximately equal widths of its components. Collisions between stable solitons were systematically studied too by means of direct simulations. The collisions are inelastic, making initially larger and smaller solitons still larger and smaller, respectively, and leading to excitation of internal oscillations in the colliding solitons. A family of two-component solitons, with two free parameters, has been found in an approximate analytical form, assuming weak ponderomotive feedback of the HF component onto the LF one. A one-parameter (i.e., non-generic) family of exact two-component solitons has been produced under condition $\gamma < 3\beta$. Perturbed dynamics of the two-component solitons, induced by a spatial shift of their HF component against the LF one, and collision between the solitons were analyzed too, by means of systematic direct simulations, with the conclusion that the shift excites a robust intrinsic mode in the solitons, and the collisions are inelastic.

All the above-mentioned species of two-component solitons are ones dominated by the LF component, and they all run with positive velocities (as *supersonic solitons* in the Davydov's model). We have also demonstrated, in the numerical form and by means of the quasi-adiabatic approximation, that the same system gives rise to HF-dominated soliton complexes, which travel with negative velocities (*subsonic solitons*). The solitons of the latter type typically appear in a combination with a quasi-linear dispersive component of the LF field.

## Acknowledgments

The work of B.A.M. is supported, in part, by the joint program in physics between NSF and Binational (US-Israel) Science Foundation through project No. 2015616, and by the Israel Science Foundation through grant No. 12876/17.

## References

1. V. E. Zakharov, S. V. Manakov, S. P. Novikov, and L. P. Pitaevskii. *Solitons: the Inverse

*Scattering Transform Method* (Nauka publishers: Moscow, 1980 (in Russian); English translation: Consultants Bureau, New York, 1984); M. Ablowitz and H. Segur, *Solitons and the Inverse Scattering Transform* (SIAM: Philadelphia, 1981); R. K. Dodd, J. C. Eilbeck, J. D. Gibbon, and H. C. Morris, *Solitons and Nonlinear Wave Equations* (Academic Press: London, 1982); A. C. Newell. *Solitons in Mathematics and Physics* (SIAM: Philadelphia, 1985); E. Infeld and G. Rowlands, *Nonlinear Waves, Solitons, and Chaos* (Cambridge University Press: Cambridge, 2000); L. A. Dickey, *Soliton Equations and Hamiltonian Systems* (World Scientific: New York, 2005).

2. R. Rajaraman, *Solitons and Instantons: An Introductions to Solitons and Instantons in Quantum Field Theory* (North Holland: Amsterdam, 1982); V. G. Baryakhtar, M. V. Chetkin, B. A. Ivanov, and S. N. Gadetskii, *Dynamics of Topological Magnetic Solitons: Experiment and Theory* (Springer-Verlag: Berlin, 1994); A.V. Ustinov, *Solitons in Josephson Junctions: Physics of Magnetic Fluxons in Superconducting Junctions and Arrays* (Wiley: Hoboken, 2005); T. Vachaspati, *Kinks and Domain Walls* (Cambridge University Press: Cambridge, 2006).

3. Y. Yang, *Solitons in Field Theory and Nonlinear Analysis* (Springer: New York, 2001).

4. G. P. Agrawal, *Nonlinear Fiber Optics* (Academic Press: San Diego, 2001); Y. S. Kivshar and G. P. Agrawal, *Optical Solitons: From Fibers to Photonic Crystals* (Academic Press: San Diego, 2003).

5. B. A. Malomed, D. Mihalache, F. Wise, and L. Torner, J. Opt. B: Quantum Semicl. Opt. **7**, R53 (2005); J. Phys. B: At. Mol. Opt. Phys. **49**, 170502 (2016); D. Mihalache, Rom. Journ. Phys. **59**, 295 (2014); B. A. Malomed, Eur. Phys. J. Special Topics **225**, 2507 (2016).

6. B. A. Malomed, *Soliton Management in Periodic Systems* (Springer: New York, 2006).

7. T. Dauxois and M. Peyrard, *Physics of Solitons* (Cambridge University Press: Cambridge, 2006).

8. L. Ostrovsky, *Asymptotic Perturbation Theory of Waves* (Imperial College Press: London, 2015).

9. V. E. Zakharov, Sov. Phys. JETP **33**, 927 (1971).

10. E.A. Ostrovskaya, S. F. Mingaleev, Yu. S. Kivshar, Yu. B. Gaididei, and P. L. Christiansen, Phys. Lett. A **282,** 157 (2001).

11. E. G. Fan, J. Phys. A: Math. Gen. **35**, 6853 (2002).

12. D. Kaya and S. M. El-Sayed, Phys. Lett. A **313**, 82 (2003).

13. P. Janssen, *The Interaction of Ocean Waves and Wind* (Cambridge University Press: Cambridge, 2009).

14. M. Brunetti, N. Marchiando, N. Berti and J. Kasparian, Phys. Lett. A **378**, 1025 (2014).

15. R. J. Wahl and W. J. Teague, J. Phys. Oceanography **13**, 2236 (1983).

16. C. Kharif, R. A. Kraenkel, M. A. Manna, and R. Thomas, J. Fluid Mech. **664**, 138 (2010).

17. W. Craig, P. Guyenne, and C. Sulem, Natural Hazards **57**, 617 (2011).

18. A. S. Davydov and N,. I. Kislukha, Physica Status Solidi B **75**, 735 (1976); A. S. Davydov, Physica Scripta **20**, 387 (1979); V. Y. Antonchenko, A. S. Davydov, and A. V. Zolotaryuk, Physica Status Solidi B **115**, 631 (1983); A. S. Davydov, *Solitons in Molecular Systems* (Reidel: Dordrecht, 1987); A. Scott, Phys. Rev. A **26**, 578 (1982); A. Scott, Physica Scipta **29**, 279 (1984); A. Scott, Phys. Rep. **217**, 1 (1992).

19. K. Hayata and M. Koshiba, Phys. Rev. Lett. **71**, 3275 (1993); A.V. Zolotaryuk, K.H. Spatschek, and A.V. Savin, Phys. Rev. B **54**, 266 (1996).

20. L. A. Cisneros-Ake and J. F. Solano Pelaez, Physica D **346**, 20 (2017).

21. E. Gromov and B. Malomed, Chaos **26**, 123118 (2016).

22. Y.B. Gaididei, P.L. Christiansen, S.F. Mingaleev, Phys. Scripta **51**, 289 (1995).

23. A. V. Zolotaryuk, K. H. Spatschek, and A. V. Savin, Phys. Rev. B **54**, 266 (1996).

23. I. V. Barashenkov and E. Y. Panova, Physica D **69**, 114 (1993); I. V. Barashenkov, Phys. Rev. Lett. **77**, 1193 (1996).

24. K. Batra, R. P. Sharma, and A. D. Verga, J. Plasma Phys. **72**, 671 (2006); F. Haas and P. K. Shukla, Phys. Rev. E **79**, 066402 (2009).

25. L. D. Landau and E. M. Lifshitz, *Quantum Mechanics* (Nauka Publishers: Moscow, 1974).